\documentclass[10pt]{article} 
\usepackage{graphicx}
\usepackage{bm}
\usepackage{amssymb}
\usepackage{amsmath,amsthm,amsfonts} 
\usepackage{caption,subfig}
\usepackage{hyperref}

\title{Semi-classical behavior of  P\"oschl-Teller coherent states}

\author{Herv\'e Bergeron$^{a}$, Jean-Pierre Gazeau$^{b}$, Petr Siegl$^{b,c,d}$, Ahmed Youssef$^{b}$}
\date{
\small
\emph{
\begin{quote}
\begin{itemize}
\item[$a)$] Universit\'e Paris-Sud, ISMO, FRE 3363, B\^at. 351 F-91405 Orsay, France \\
Email: herve.bergeron@u-psud.fr
\item[$b)$] Paris Diderot Paris 7, Laboratoire APC, Case 7020, 75205 Paris Cedex 13, France\\
Email: gazeau@apc.univ-paris7.fr, ahmed.youssef@apc.univ-paris7.fr
\item[$c)$] Czech Technical University in Prague, FNSPE, B\v rehov\'a 7, 11519 Prague,\\  Czech Republic 
\item[$d)$] Nuclear Physics Institute ASCR, 25068 \v Re\v z, Czech Republic \\
Email: siegl@ujf.cas.cz
\end{itemize}
\end{quote}
}
\medskip
\today
}

\def\E{\mathcal{E}}

\def\eqdef{\stackrel{\mbox{\tiny def}}{=}}     
\newcommand{\ket}[1]{|\kern.3ex#1\kern.3ex\rangle}
\newcommand{\bra}[1]{\langle\kern.3ex #1 \kern.3ex|}
\newcommand{\scalar}[2]{\langle\kern.3ex #1 \kern.3ex|\kern.3ex#2\kern.3ex\rangle}

\begin{document}
\maketitle

\begin{abstract}
We present a construction of semi-classical states for P\"oschl-Teller
potentials based on a  supersymmetric quantum mechanics approach. The parameters of these ``coherent" states are points in the classical phase space of these systems. They minimize a special uncertainty relation. Like standard coherent states they resolve the identity with a uniform measure. They permit to establish the correspondence (quantization) between classical and quantum quantities. Finally, their   time evolution is localized on the classical phase space trajectory.
\end{abstract}

\section{Introduction}
The search of quantum states that exhibit a semi-classical time behavior was initiated 
in Schr\"odinger's pioneering work on ``packets of eigenmodes" of the harmonic oscillator \cite{Schrodinger-1926-14}.

\begin{quotation}
\emph{I show that the packet of eigenmodes with high quantum number $n$ and with relatively small difference in quantum numbers may represent the mass point that moves according to the usual classical mechanics, i.e. it oscillates with [classical] frequency  $\nu_0$.}
\end{quotation}
These Schr\"odinger states have been qualified as coherent by Glauber within the context of quantum optics. 
With regard to the huge amount of recent works on quantum dots and quantum wells in nanophysics 
it has become challenging to construct quantum states for infinite wells which display  localization properties comparable to those nicely displayed by the  Schr\"odinger states. Infinite wells are often modeled by P\"oschl-Teller (also known as trigonometric Rosen-Morse)  confining potentials \cite{Poschl-1933-83,Rosen-1932-42} used, e.g., in quantum optics \cite{Yildirim-2005-72, Wang-2007-75}. The infinite square well is a limit case of this family  referred to in what follows as  \mbox{${\mathcal{T}}$-potentials}. The question  is to find  a family of states: a) phase-space labelled, b) yielding a resolution of the identity with respect to the usual uniform measure, and  c) exhibiting semi-classical  phase space properties with respect to ${\mathcal{T}}$-Hamiltonian time evolution. We refer to these states as coherent states (CS) as they share many striking properties with Schr\"odinger's original semi-classical states. 

${\mathcal{T}}$-potentials belong to the class of shape invariant potentials \cite{Gendenshtein-1983-38} intensively studied within the framework of supersymmetric quantum mechanics (SUSYQM) \cite{Cooper-2002}. Various semi-classical states adapted to ${\mathcal{T}}$-potentials have been proposed  in previous  works \cite{ Aleixo-2007-40,Antoine-2001-42,Crawford-1998-57,Shreecharan-2004-69} and references therein. However, they do not verify simultaneously a), b), and c). Moreover, correspondence  between  classical and quantum  momenta requires a  thorough analysis since there exists  a well-known ambiguity  in the definition of  a quantum momentum operator \cite{Reed-1980, Antoine-2001-42}. This is due to the confinement of the system in an interval, unlike the harmonic oscillator case.

In this letter, we present a construction of coherent states for ${\mathcal{T}}$-potentials based on a general approach given  by one of us in \cite{Bergeron-1995-36}. We examine in detail  the classical-quantum correspondence based on these states (``CS quantization"). We eventually show that  our states  stand comparison with the Schr\"odinger CS in terms of semi-classical time behavior.

\section{Definition of SUSYQM \mbox{coherent states}}
Let us consider the motion of a  particle confined in the interval $[0,L]$ and submitted to  the repulsive symmetric \mbox{${\mathcal{T}}$-potential} 
 \begin{equation}
 V_{\nu}(x)=\E_0 \frac{ \: \nu(\nu+1)}{\sin^2 \frac{\pi}{L}x}\, ,
 \end{equation}
where $\nu \ge 0$ is a dimensionless parameter.  The limit \mbox{$\nu \to 0$} corresponds to the infinite square well. The factor   $\E_0 = \hbar^2 \pi^2 (2 m L^2)^{-1} \geq 0$ is chosen as the ground state energy of the infinite square well. On the quantum level, the  Hamiltonian acts in the Hilbert space $\mathcal{H}=L^2([0,L],dx)$ as:
\begin{equation}
{\bf H}_{\nu}=-\frac{\hbar^2}{2m}\frac{d^2}{dx^2}+V_{\nu}(x)\, .
\end{equation}

The eigenvalues  $E_{n,\nu}$ and corresponding eigenstates $\ket{\phi_{n,\nu}}$ of ${\bf H}_{\nu}$ read
\begin{equation}
 E_{n}=\E_0(n+\nu+1)^2, \ \ \ n=0,1,2..., 
 \end{equation}
 \begin{equation}
\phi_{n}(x)= Z_{n} \sin^{\nu+1}\left(\frac{\pi}{L}x\right)\,  \mathrm{C}_n^{\nu+1}\left(\cos \frac{\pi}{L}x\right)
 \end{equation}
where $C_n^{\nu+1}$ is  a  Gegenbauer polynomial and
 \begin{equation}
 Z_{n}=\Gamma(\nu+1) \frac{2^{\nu+1/2}}{\sqrt{L}} \sqrt{\frac{n! (n+\nu+1)}{\Gamma(n+2\nu+2)}}
 \end{equation}
is the normalization constant. Eigenfunctions $\phi_{n}$ obey the Dirichlet boundary conditions $\phi_{n}(0)=\phi_{n}(L)=0$. A detailed mathematical discussion on the boundary conditions and self-adjoint extensions for the ${\mathcal{T}}$-Hamiltonian can be found in \cite{Gesztesy-1985-362,Antoine-2001-42}. 

In particular, the ground state eigenfunction $\phi_{0}$ is $Z_{0} \sin^{\nu+1} \frac{\pi}{L}x$
and the eigenfunctions for the infinite square well ($\nu=0$) reduce to $ \sqrt{\frac{2}{L}} \sin \frac{(n+1)\pi}{L} x.$

We define the superpotential $W_{\nu}(x)$ as 
\begin{equation}
W_{\nu}(x)\eqdef-\hbar \frac{\phi'_{0}(x)}{\phi_{0}(x)}=-\frac{\hbar \pi}{L} (\nu+1) \cot \frac{\pi}{L}x
\end{equation}
and the lowering and raising operators ${\bf A}_{\nu}$ and ${\bf A}^\dag_{\nu}$ as
\begin{equation}
{\bf A}_{\nu} \eqdef W_{\nu}(x)+\hbar \frac{d}{dx} \text{ and } {\bf A}^\dag_{\nu} \eqdef W_{\nu}(x)-\hbar \frac{d}{dx}
\end{equation}
Thus, the ${\mathcal{T}}$-Hamiltonian ${\bf H}_{\nu}$ can be rewritten in terms of these operators:
\begin{equation}
{\bf H}_{\nu}=\frac{1}{2m}{\bf A}^\dag_{\nu} {\bf A}_{\nu} +E_0. 
\end{equation}
As expected \cite{Cooper-2002}, the supersymmetric partner ${\bf H}^{(S)}_{\nu}$ 
\begin{equation}
{\bf H}^{(S)}_{\nu}=\frac{1}{2m} {\bf A}_{\nu} {\bf A}^\dag_{\nu} + E_0. 
\end{equation}
coincides with the original Hamiltonian with increased $\nu$: ${\bf H}^{(S)}_{\nu}= {\bf H}_{\nu+1}.$

The classical phase space for the motion in a \mbox{${\mathcal{T}}$-potential} is defined as the infinite band in the plane: 
\mbox{$
\mathcal{K}=\{ (q,p) | q\in [0,L] \text{ and } p\in \mathbb{R} \}\, .
$}
Let us introduce the operators 
\begin{equation}
\label{QP}
{\bf Q}: \psi(x) \mapsto x\psi(x), \quad  \textrm{and} \quad {\bf P}: \psi \mapsto -i \hbar  \dfrac{d}{dx} \psi(x).
\end{equation}
We then build  our  coherent states $|\eta_{q,p}\rangle$ as  normalized eigenvectors of ${\bf A}_{\nu} =W_{\nu}({\bf Q})+i {\bf P}$ with eigenvalue \mbox{$W_{\nu}( q)+i p$}, the latter being  the classical counterpart of $A_\nu$ as shown below (see Table \ref{table1}).
\begin{equation}
\label{CSPT}
\ket{\eta_{q,p}}= N_{\nu}(q) \left\vert\xi_{W(q)+ip}^{[\nu]}\right\rangle\,, \ (q,p)\in \mathcal{K}\, , 
\end{equation}
where $\xi_z(x)= e^{z x/\hbar} \sin^{\nu+1} \left(\frac{\pi}{L}x \right)\text{ for } x \in [0,L]\, . $
The normalization coefficient $N_{\nu}(q)$ is given by
\begin{align}
\nonumber \frac{1}{N_{\nu}^2(q)}=&\frac{2^{\nu+1} |\Gamma(\nu+2-i (\nu+1) \cot \frac{\pi}{L}q)|}{\sqrt{L} \sqrt{ \Gamma(2\nu+3)}}
\exp\left[\frac{\pi}{2}(\nu+1)\cot \frac{\pi}{L}q \right].
\end{align}
For the sake of simplicity we drop off systematically in the sequel the $\nu$ dependence of various used symbols when no confusion is possible. 
It is possible to show that the function $x \mapsto |\eta_{q,p}(x)|$ reaches its maximal value for $x=q$ and
$\langle \mathbf{P} \rangle_{p,q}=p$. Finally, the uncertainty relation $\Delta W_\nu(\mathbf{Q}) \Delta \mathbf{P} \geq \frac{\hbar}{2} \langle W_\nu'(\mathbf{Q}) \rangle$ is minimized by our CS as proved in \cite{Bergeron-1995-36}.

\section{CS Quantization and \mbox{expected values}}
As is proved in \cite{Bergeron-new}, the CS family \eqref{CSPT} resolves the unity with respect to the  uniform measure on the phase space $\mathcal{K}$:
\begin{equation}
\label{equa:resol}
\int_{\mathcal{K}} \frac{dq\,dp}{2 \pi \hbar} \ket{\eta_{q,p}} \bra{\eta_{q,p}} = \mathbb{I}\, .
\end{equation}
As an immediate consequence we proceed with the CS quantization of ``classical observables"  $f(q,p)$ through  the correspondence \cite{Klauder-1985,Gazeau-2009}
\begin{equation}
\label{quantization}
f(q,p) \to {\bf F}=\int_{\mathcal{K}} \frac{dq\, dp}{2 \pi \hbar} f(q,p) \ket{\eta_{q,p}}\bra{\eta_{q,p}}\, .
\end{equation}
This operator-valued integral is  understood  as the sesquilinear form, 
\begin{equation}
B_f(\psi_1,\psi_2)=\int_{\mathcal{K}} \frac{dq\, dp}{2 \pi \hbar} f(q,p) \scalar{\psi_1}{\eta_{q,p}}\scalar{\eta_{q,p}}{\psi_2}.
\end{equation}
The form $B_f$ is  assumed to be  defined on a dense subspace of the Hilbert space.  If $f$ is real and at least semi-bounded, the Friedrich's extension \cite[Thm. X.23]{Reed-1975} of $B_f$ univocally defines a self-adjoint operator. However, if $f$ is not semi-bounded, there is no natural choice of a self-adjoint operator associated with $B_f$. In this case, we can consider directly the symmetric operator $\mathbf{F}$ given by Eq.\eqref{quantization} enabling us to obtain a self-adjoint extension (unique for particular operators). The question of what is the class of operators that may be so represented is a subtle one \cite{Klauder-1985,Gazeau-2009}.
In Table \ref{table1}, we give a list of  operators obtained through  the CS quantization of basic functions $f$. One can also compute the so-called ``lower'' or ``covariant''  symbols \cite{Klauder-1985,Gazeau-2009} of operators defined as the expectation values of the latter  in the CS. 
In Table \ref{table2} we give a list of  functions of the most important quantum operators.

\begin{table}[htb!]
\renewcommand{\tabcolsep}{0.1cm}
  \centering 
\newcommand\T{\rule{0pt}{2.8ex}}
\newcommand\B{\rule[-1.5ex]{0pt}{0pt}}
\begin{tabular}{|c|c|c|c|c|}

\hline
Name  \T & $f$ & $\mathbf{A}_f$ & Operator action  & Properties  \B\\
\hline \hline
 Position \T & $q$ & $ F(\mathbf{Q}) $ (*)& multiplication    &bounded   \\
 & & &   &self-adjoint  \B \\
 \hline
 Superpotential \T  & $W_\nu(q)$ & $W_\nu(\mathbf{ Q})$& multiplication  & unbounded \\
 & & & & self-adjoint  \B \\
   \hline
 Potential \T  & $\frac{1}{\sin^2 \pi q/L}$&$\frac{(2\nu+3)(2\nu+2)^{-1}}{\sin^2 \pi \mathbf{ Q}/L}$ & multiplication   &unbounded \\
 & & & & self-adjoint  \B \\
   \hline
``Momentum" \T   & p & $\mathbf{ P}$ & $\mathbf{P} \phi_n=-i \hbar \phi'_n $  & unbounded \\
& & &  & symmetric   \B \\
   \hline
   Hamiltonian \T   & $ \frac{p^2}{2m}+ \frac{2\nu-1}{2\nu+3} \frac{\E_0(\nu+1)^2}{\sin^2 \pi q/L}$ & $\mathbf{ H}_\nu$  & Schr\"odinger &  semi-bounded \\
&  & &operator  &   self-adjoint\\
&  & & &  ($\nu \geq 1/2$)   \B \\
   \hline
\end{tabular}
  \caption{Some quantized classical observables. 
The operators $\mathbf{Q}$ and $\mathbf{P}$ are defined in eq. \eqref{QP}.
(*)$F(x)= \sin^{2\nu+2} (\pi x/L) \int_0^L dq\: q N_\nu^2(q) \exp (2 W_\nu(q) x/L)
$.}
\label{table1}
\end{table}

\begin{table}[htb!]
\newcommand\T{\rule{0pt}{2.8ex}}
\newcommand\B{\rule[-1.5ex]{0pt}{0pt}}
  \centering 

\begin{tabular}{|c|c|c|}
\hline
Name  \T & $\mathbf{A}$ & $f$  \B \\
\hline
\hline
Position \T & $\mathbf{Q}$& $N_\nu^2(q)$ \\
 & & \B $\times\int_0^L dx \:x \sin^{2\nu+2} \frac{\pi x}{L}  e^{\frac{2 W_\nu(q) x}{L}} $ \\
\hline
``Momentum" (*) \T & $\mathbf{P}$& $p$ \B \\
\hline
Superpotential \T  &  $W_\nu(\mathbf{Q})$ & $W_\nu(q)$ \B \\ 
\hline
Potential \T & $\frac{1}{\sin^2 \pi \mathbf{Q}/ L}$ & $\frac{2\nu+2}{2\nu+1}\frac{1}{\sin^2 \pi q/ L}$ \B \\
\hline
Kinetic energy \T &$\frac{\mathbf{P}^2}{2m}$& $\frac{p^2}{2m}+ \frac{1}{2\nu+1} \frac{\E_0 (\nu+1)^2}{\sin^2 \pi q/L}$ \B \\
\hline
\end{tabular}
  \caption{Some lower symbols. (*) The operator $\mathbf{P}$ is the one given in Table \ref{table1}.}\label{table2}
\end{table}

\section{Semi-classical behavior}
\begin{figure}[!ht]
    \centering
\subfloat
[Phase space distribution \eqref{eqn:phasespacerepre} for $\nu=0$ of the state $\eta_{q_0,p_0}$ with $q_0=L/5$, 
$p_0=4 \pi \hbar/L$ and $L=20\,$\AA. The thick curve is the expected phase trajectory in the infinite square well, 
deduced from the semi-classical hamiltonian in Eq. \eqref{eqn:semiclassenergy}. The particle is an electron, its mean energy deduced from Eq. \eqref{eqn:semiclassenergy} is $E=1.6 \: \mathrm{eV}$. Increasing values of the function are encoded by the colors from blue to red.  Note that for Schr\"odinger CS the corresponding distribution is a Gaussian localized on points  of  a circular trajectory in the complex plane. See \href{http://gemma.ujf.cas.cz/~siegl/SUSYCS.html}{animation} for an animated time-evolution.]
{\includegraphics[width =0.4\textwidth]{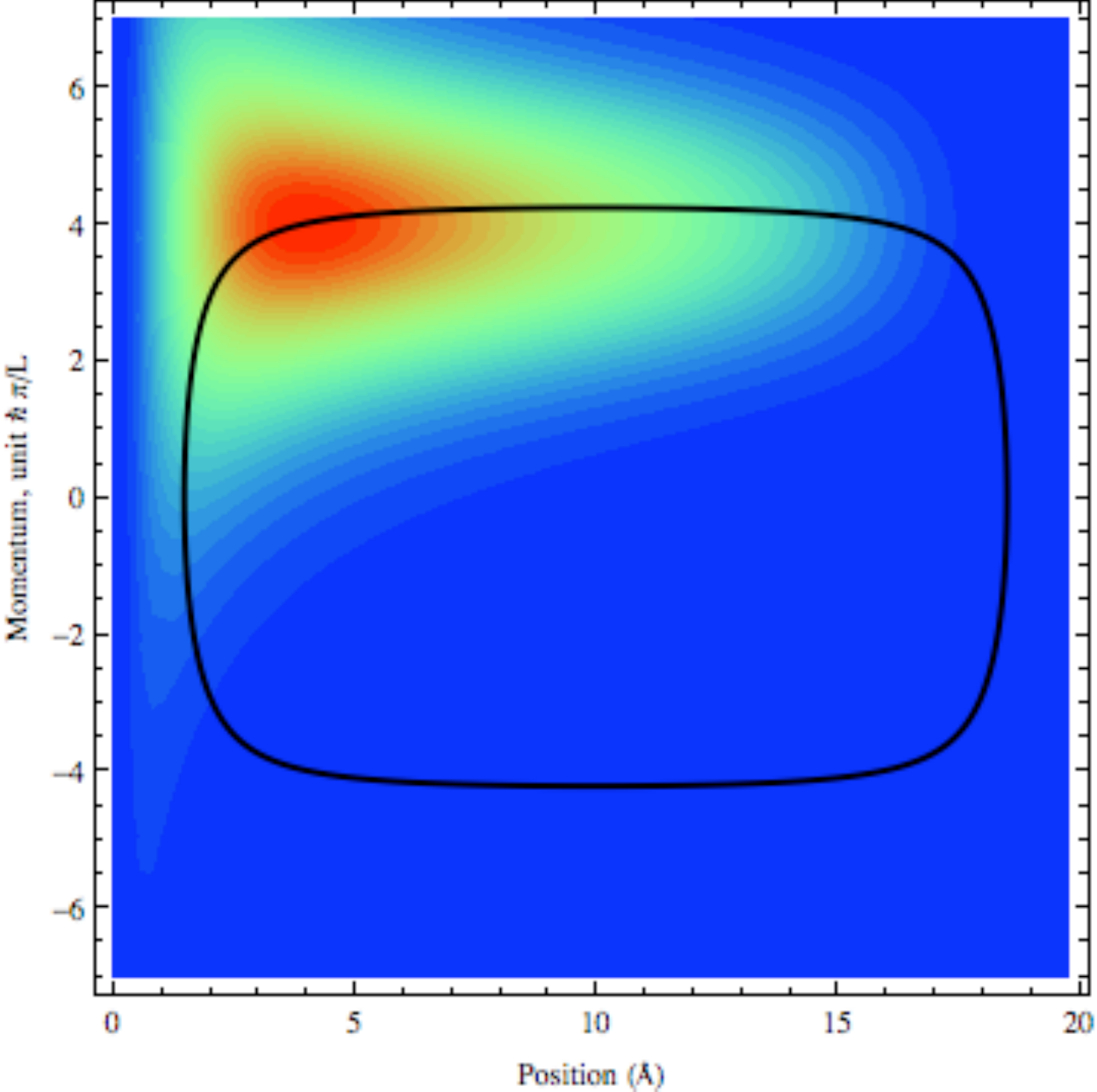}} \hspace{0.5cm} 
\subfloat
[Time average of the phase space distribution of $\eta_{q_0,p_0}(t)$ evolving following the Hamiltonian of the infinite square well. The values of parameters and the thick curve are those of Figure \ref{figure}$(a)$. Increasing values of the function are encoded by the colors from blue to red. Note that for Schr\"odinger CS, the corresponding time-averaged distribution would be uniformly localized on a circular trajectory in the complex plane.]
{\includegraphics[width =0.4\textwidth]{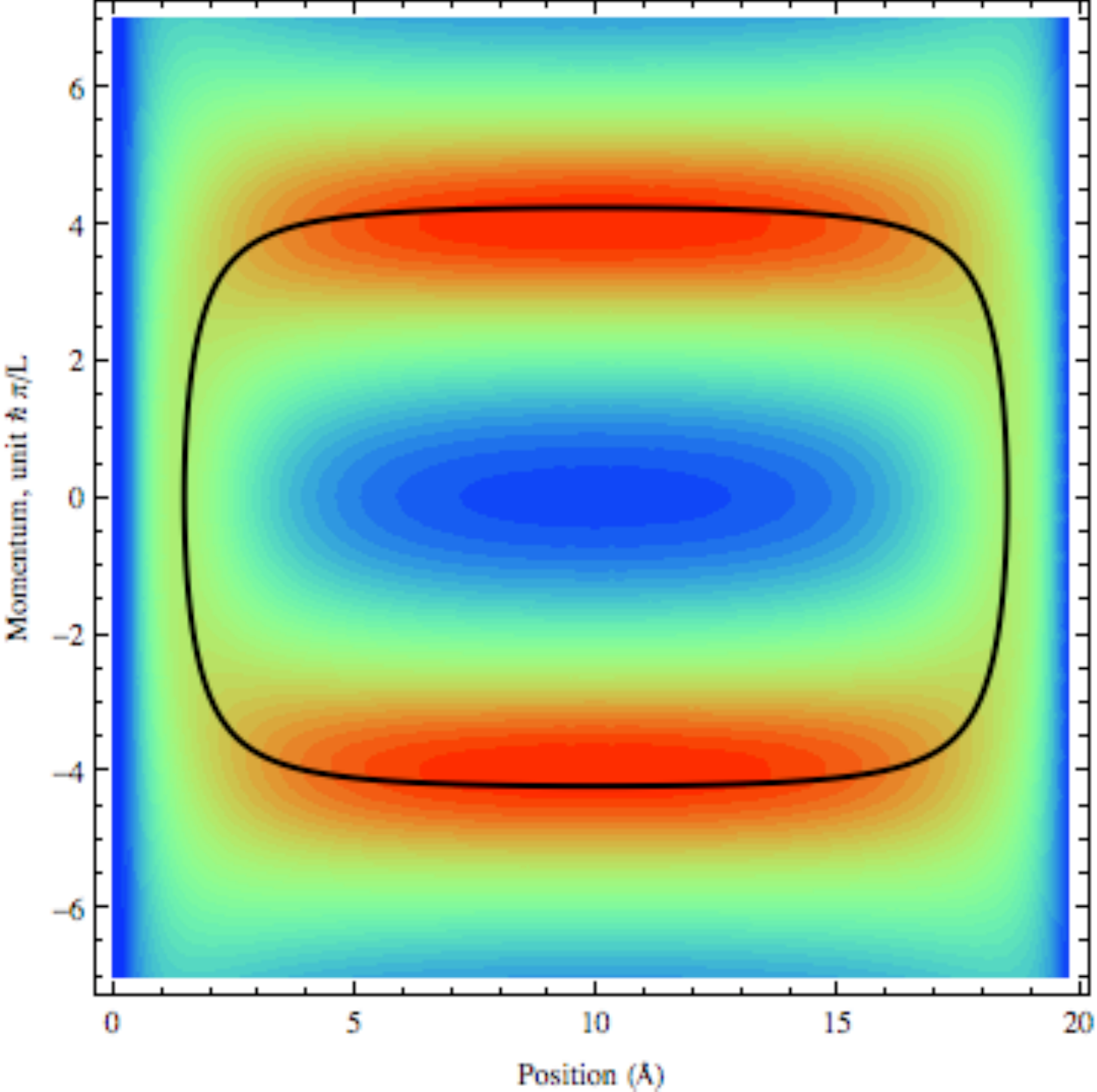}}
    \caption{Phase space distribution of CS and its time average }
    \label{figure}
\end{figure}

For any normalized state $\phi \in \mathcal{H}=L^2([0,L],dx)$, the resolution of unity \eqref{equa:resol}  allows us to get its phase space representation \mbox{$\Phi(q,p)  \eqdef \scalar{\eta_{q,p}}{\phi}/\sqrt{2\pi\hbar}$} and the resulting   probability distribution on the phase space $\mathcal{K}$:
\begin{equation}
\label{eqn:phasespacerepre}
\mathcal{K} \ni (q,p)\mapsto  \frac{1}{2\pi \hbar} |\scalar{\eta_{q,p}}{\phi}|^2=\rho_{\phi}(q,p)
\end{equation}
The  phase-space distribution $\rho_{\eta_{q_0,p_0}}(q,p)$ for a particular state $\phi=|\eta_{q_0,p_0}\rangle$ is shown in Figure \ref{figure}$(a)$ for $\nu=0$ (infinite square well), see \href{http://gemma.ujf.cas.cz/~siegl/SUSYCS.html}{animation} for an animated time-evolution.

%

Let us now examine the time behavior $t \mapsto \rho_{\phi(t)}(q,p)$ for a state $\phi(t)$ evolving under the action of the infinite square well Hamiltonian ${\bf H}_0$ (there is no significant difference from a generic $\nu \neq 0$ case):
\begin{equation}
\ket{\phi(t)}=e^{-i {\bf H_0} t/\hbar} \ket{\phi}=\sum_{n=0}^\infty e^{-i \E_0 (n+1)^2 t/ \hbar} \scalar{\phi_{n,0}}{\phi} \ket{\phi_{n,0}}
\end{equation}
where $\phi_{n,0}\equiv  \sqrt{\frac{2}{L}} \sin \frac{(n+1)\pi}{L} x$.

With  $\phi=\eta_{q_0,p_0}$ as an initial state, we have  for a given $\nu$ (see Table \ref{table2})
\begin{equation}
\label{eqn:semiclassenergy}
\scalar{\eta_{q_0,p_0}(t)}{{\bf H}_0 |\eta_{q_0,p_0}(t)} =\frac{p_0^2}{2m}+ \frac{1}{2\nu+1} \frac{\E_0(\nu+1)^2}{\sin^2 \frac{\pi}{L} q_0}.
\end{equation} 
Since the lower symbols of $W_\nu(\mathbf{Q})$ and $\mathbf{P}$ correspond to their classical original functions $W_\nu(q)$ and $p$ (see Table \ref{table2}), one can expect that the time average of the probability law $\rho_{\eta_{q_0,p_0}(t)}(p,q)$ corresponds to some fuzzy extension in phase space of the classical trajectory corresponding to the time-independent Hamiltonian in the r.h.s. of \eqref{eqn:semiclassenergy}.
This key result is illustrated in Figure \ref{figure}(b) where we have represented the time average  distribution $\bar{\rho}$ defined as
\begin{equation}
\label{rhobar}
\bar{\rho}(q,p)=\lim_{T \to \infty} \frac{1}{T} \int_0^T \rho_{\eta_{q_0,p_0}(t)}(q,p) dt,
\end{equation}
for the same values of the parameters as in Figure \ref{figure}$(a)$. 
The time average distribution $\bar{\rho}$ allows us to compare the quantum behavior with the classical trajectory, but the expression \eqref{rhobar} hides the complex details of the wave-packet dynamics. The latter exhibits a splitting of the initial wave-packet into secondary ones during the sharp reflection phase, each of them following the classical trajectory, before they amalgamate to reconstitute a unique packet (revival time). This important point makes the difference with the time behavior of the Schr\"odinger states for the harmonic oscillator.

\section{Conclusion}
We have presented  a family of CS for the ${\mathcal{T}}$-potentials that sets a sort of natural bridge between  the phase space and its quantum counterpart. These CS share with the Schr\"odinger ones some of their most striking properties, e.g.  resolution of identity with uniform measure and saturation of uncertainty inequalities. They also possess remarkable evolution stability features (not to be confused with CS temporal stability in the sense of \cite{Antoine-2001-42} corresponding to the time parametric evolution): their time evolution generated by $\mathbf{H}_\nu$  is localized on the classical phase space trajectory. 
The  approach developed in this paper  can be easily extended to higher dimensional  bounded  domains, provided that the latter be symmetric enough (e.g. square, equilateral triangle, etc) to allow shape invariance integrability.


\section*{Acknowledgments}
P.S. appreciates the support by the Grant Agency of the Czech Republic
project No. 202/08/H072 and the Czech Ministry of Education, Youth and
Sports within the project LC06002. We wish to thank A. Comtet, J. Dittrich, P. Exner, J. R. Klauder, T. Paul and J. Tolar  for fruitful discussions and comments. 

{\footnotesize
\bibliographystyle{unsrt}
\bibliography{C:/Data/00Synchronized/GlobalReferences}
}

\end{document}